# Anomalous and Topological Hall Resistivity in Ta/CoFeB/MgO Magnetic Systems for Neuromorphic Computing Applications

Aijaz H. Lone, Xuecui Zou, Gianluca Setti and Hossein Fariborzi

*Computer, Electrical and Mathematical Science and Engineering Division,*
*King Abdullah University of Science and Technology, Thuwal, Saudi Arabia*

Debasis Das, Xuanyao Fong

*Department of Electrical and Computer Engineering,*
*National University of Singapore, Singapore 117583*

**Topologically protected spin textures, such as magnetic skyrmions, have the potential for dense data storage as well as energy-efficient computing due to their small size and a low driving current. The evaluation of the writing and reading of the skyrmion's magnetic and electrical characteristics is a key step toward the implementation of these devices. In this paper, we present the magnetic heterostructure Hall bar device and study the anomalous Hall and topological Hall signals in the device. Using the combination of different measurements like magnetometry at different temperatures, Hall effect measurement from 2K to 300K, and magnetic force microscopy imaging, we investigate the magnetic and electrical characteristics of the magnetic structure. We measure the skyrmion topological resistivity at different temperatures as a function of the magnetic field. The topological resistivity is maximum around the zero magnetic field and it decreases to zero at the saturating field. This is further supported by MFM imaging. Interestingly the resistivity decreases linearly with the field, matching the behavior observed in the corresponding micromagnetic simulations. We combine the experimental results with micromagnetic simulations, thus propose a skyrmion-based synaptic device and show spin-orbit torque-controlled potentiation/depression in the device. The device performance as the synapse for neuromorphic computing is further evaluated in a convolutional neural network CNN. The neural network is trained and tested on the MNIST data set we show devices acting as synapses achieving a recognition accuracy close to 90%, on par with the ideal software-based weights which offer an accuracy of 92%.**

## I. INTRODUCTION

Magnetic skyrmions have been investigated for their applications in non-volatile data storage and computing applications[1,2,3,4]. These swirling spin structures have the advantage of being topologically protected against external perturbations[5,6,7]. Skyrmions are stabilized via the chiral Dzyaloshinskii–Moriya interaction (DMI) interactions in non-centrosymmetric magnetic compounds[8,9] and due to interface DMI originating in the thin film systems exhibiting broken inversion symmetry[10,11,12]. This chiral DMI antisymmetric exchange interaction responsible for forming magnetic skyrmions[13,14], emerges out of the strongspin–orbit coupling at a heavy metal/ferromagnetic (HM/FM) interfaces with broken inversion symmetry[15,16,17]. The exchange coupling energy and anisotropy energy prefer the parallel and uniaxial alignment of spins, but the DMI and dipolar energy terms prefer the noncollinear alignment of spins[18,19]. In asymmetric ferromagnetic multilayer systems, such as Pt/Co/Ta [20], Pt/CoFeB/MgO [12], and Ta/CoFeB/MgO[10] the DMI is obtained by the high interfacial spin–orbit coupling caused by symmetric breaking. Along with DMI recent reports have shown that dipolar interactions can also stabilize the skyrmions in a thin film system[21,22,23]. Skyrmionics as a sub-field of spintronics has gained substantial attention as the magnetic skyrmions promise increased memory density owing to the scalable [24,25] and sub-nanometric skyrmion size[26,27]. Skyrmions can be stabilized at room temperature and driven by current densities ~ $1 \times 10^4 A/cm^2$ [28,24] so writing into these devices is energy-efficient when compared to its counterpart domain wall devices[29,30,31,32]. The intrinsic topological protection of the skyrmion gives it a lifetime and makes them stable against external perturbations [33,5]. Topological protection means that skyrmions have a characteristic topological integer that can't be changed by the continuous deformation of the field [34]. Mathematically the topological integer also known as skyrmion charge or winding number is given by [35]

$$Q = \frac{1}{4\pi} \int \int \boldsymbol{m} \cdot \left(\frac{\partial \boldsymbol{m}}{\partial x} \times \frac{\partial \boldsymbol{m}}{\partial y}\right) dx dy. \qquad (1)$$

The spins projected on the xy-plane and normalized magnetization vector $\boldsymbol{m}$ can be determined by the radial function $\theta$, Vorticity $Q_v$ and helicity $Q_h$:

$$m(r) = [\sin(\theta)\cos(Q_v\varphi + Q_h), \sin(\theta)\sin(Q_v\varphi + Q_h), \cos(\theta)]. \qquad (2)$$

The vorticity number is related to the skyrmion number as follows [4]:

$$Q = \frac{Q_v}{2}\left[\lim_{r \to \infty} \cos(\theta(r)) - \cos(\theta(0))\right]. \qquad (3)$$

The topological characteristics or non-trivial geometry of the skyrmions gives birth to an emergent electromagnetic field (EEMF)[36] that leads the other physical effects such as the topological Hall effect[37]. The conduction electrons collect a Berry phase when they adiabatically follow the skyrmion magnetic texture thus apart from the normal Hall and anomalous Hall effect[38,39]. The topological Hall effect is also expected in these magnetic thin films. Thus, by evaluating the topological Hall effect in terms of extra contribution in the Hall signal, the reading of the skyrmions can be realized. Recently there have been a few reports showing the contribution of the skyrmion Hall resistivity in MnSi[8,40], [Pt/Co/Ir]×10 [41] and 2D materials such as $Cr_2Te_3$ and $Cr_2Se_3$ [42]. But, considering the development of CoFeB/MgO material combination for spintronic devices[43,44,45,46] and especially skyrmionic devices[47,48,49,50] it becomes quite imperative to evaluate and study the magnetic characteristics and electrical characteristics of the skyrmions.

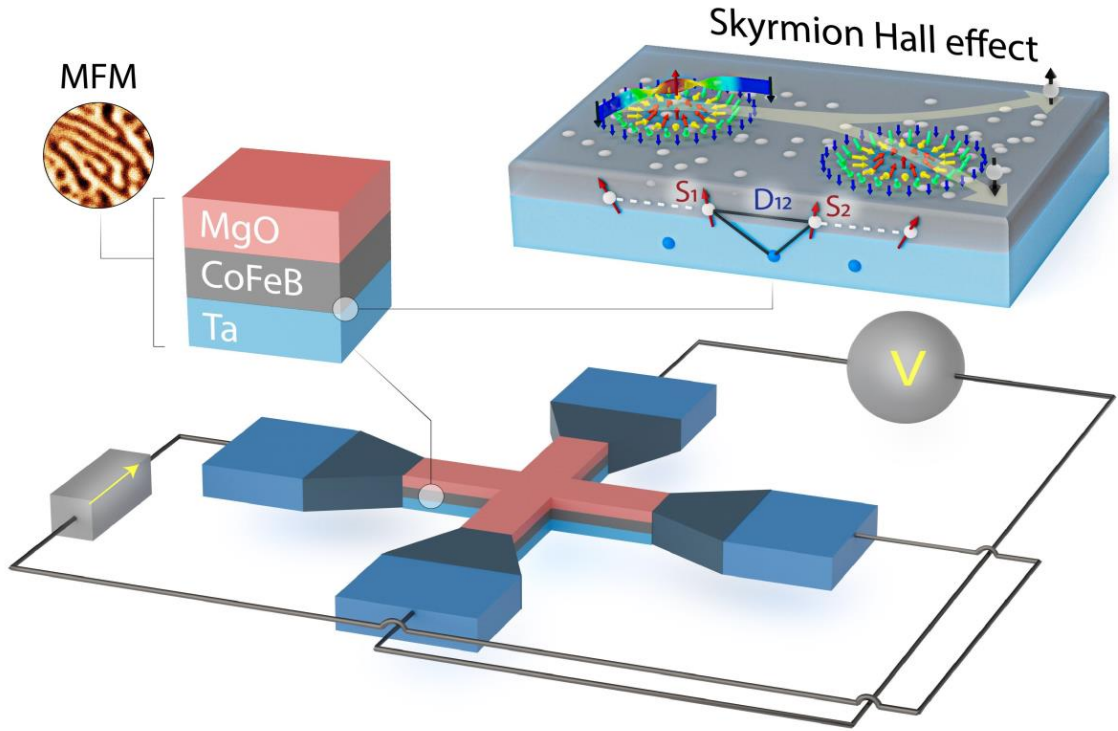

FIG. 1. Magnetic multilayer device and the setup for Anomalous Hall measurements

In this article, we present experimental and micromagnetic studies for the evaluation of the skyrmion topological resistivity in the Ta/CoFeB/MgO thin film system. We optimize the CoFeB thickness to obtain skyrmions at room temperature. Using the combination of different measurements like magnetometry at different temperatures, Hall measurement from 2K to 300K, and magnetic force microscopy imaging, we investigate the magnetic and electrical characteristics of the skyrmions. We measure the skyrmion topological resistivity at different temperatures as a function of the magnetic field. The topological resistivity is maximum around the zero fields and it decreases to zero at the saturating field. This is further supported by MFM imaging in different fields. Interestingly the resistivity decreases quite linearly with the field, which is also seen in the corresponding micromagnetic simulations. Motivated by the experimental results from anomalous and topological resistance and supported with the micromagnetic simulations we propose a scaled version of the device as the co- Domain wall/ skyrmion synapse with high linearity and show MNIST data classification performed by the neural network -based on the proposed synaptic device.

## Experimental Results and Discussion

The magnetic thin film system considered in this work is [Ta/CoFeB/MgO] $_{\times 15}$ and the device for characterization and measurements is fabricated as a crossbar of size $50 \times 2.5$ $\mu m^2$ as shown in Fig. 1. The details of the stack deposition and device fabrication can be found in the supplementary material (Fig. S1). We considered 15 repetitions to stabilize small skyrmions at room temperature and zero fields for different CoFeB thicknesses we observe skyrmions of size around 400 nm. Sample exhibits admix of stripe domains, labyrinth domains, and skyrmions of size around 300 nm. The stack deposition quality was from the cross-sectional transmission electron microscopy (TEM) of the samples and is shown in supplementary Fig. [S1-B], where we observe Ta thickness around 5 nm, CoFeB thickness around 0.85 nm, and MgO 2 nm.

Fig. 2(a) shows the magnetization-magnetic field (M-H) hysteresis loops measured using VSM for different samples with varying CoFeB thickness. The M-H loops show perpendicular magnetic anisotropy (PMA) in all the samples with thicknesses (of 0.5 nm to 1.1 nm). The butterfly shape which is characteristic of the multi-domain magnetic texture stabilization in the FM thin film is obtained in the majority of the samples. As observed the multi-domain characteristics are shown for the CoFeB thickness ranging from 0.77 nm to 1.1 nm. The saturation magnetization increases linearly with increasing thickness as shown in Fig. 2(a). The switching field obtained from the VSM measurements is reduced for all the samples with more prevalent butterfly hysteresis this indicates the magnetization switching by gradual multi-domain reversal. We observe that the CoFeB thickness ranging from 0.83 nm to 1.1 nm has a higher probability of stabilizing the skyrmions at room temperature. In Fig. 2(b) the magnetic texture was obtained from MFM imaging for the sample with a thickness 1 nm at zero field. The mixed phase with skyrmions of $Q = \pm 1$ and stripe domains is visible as expected from the VSM measurements. Corresponding MFM images of the samples with varying CoFeB thickness in zero field and at room temperature are shown in Fig. 2(c-e). The MFM probed magnetic texture of these samples reveals reducing stripe domain width with reducing CoFeB thickness from 0.95 nm down to 0.9 nm and 0.8 nm which can be attributed to the increased surface anisotropy. For CoFeB thickness around 0.9 nm, the majority of the stripe domains collapse into skyrmion bubbles of diameter around 300 nm as shown in Fig. 2(c). The saturation magnetization ($M_s$) of the $t_F$=0.9 nm is 0.73 emu/cm$^3$ which is greater than the $M_s$ for $t_F$ =1 nm so in the case of 0.9 nm the stray field is helping in skyrmion stability.

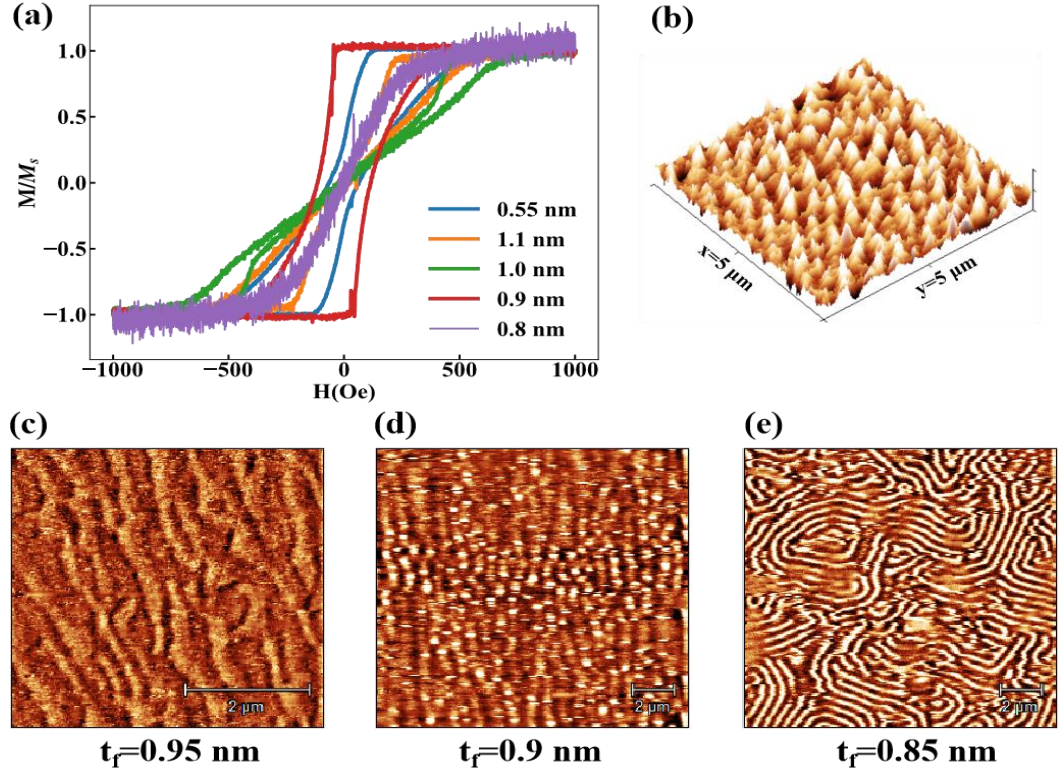

FIG. 2. a) VSM magnetic hysteresis shows butterfly shape (characteristics of multidomain/skyrmions) for magnetic layers thickness (0.55 nm to 1.1 nm). (b) MFM picture of the sample with CoFeB thickness 0.95 nm shows room temperature skyrmions. MFM probed the magnetic texture of the samples at room temperature and zero field for thickness 0.9 nm (d), 0.95 nm (c), and 0.85 nm (e) showing different magnetic phases. The width of the stripe domains decreases on reducing thickness which indicates increased surface anisotropy. For CoFeB thickness around 0.9 nm, the majority of the stripe domains collapse into skyrmion bubbles of diameter around 400 nm

## III. ANOMALOUS HALL RESISTANCE MEASUREMENTS

The sample exhibiting skyrmions with CoFeB thickness around 0.9 nm was patterned into $50 \times 2.5$ $\mu m^2$ Hall bars as shown in Fig. 3. The Hall measurements were carried out in presence of a DC 500 $\mu$A and magnetic field was swept in the range of ±10 KOe. Variations of anomalous Hall resistance ($R_{xy}$) with the applied magnetic field (H) are shown in Fig. 3 (a)-(b/c) for various temperatures ranging from 2-300 K. For temperature T= 300K, 250K, and 200K, we observe the butterfly loop behavior in $R_{xy}$ vs. H. These results are in-line with the VSM based MH loops as shown in Fig. 2. Thus, the resistance behavior observed in the Hall measurements is attributed to the multi-domain (skyrmion) switching as seen in Fig. 3(d) for T=300K the switching is more gradual and linear like magnetization switching observed in VSM. In magnetic materials, the Hall conductivity observed across transverse (*x-y*)-terminals has a contribution from the normal Hall effect due to the magnetic field, anomalous Hall effect due to the magnetization, and topological Hall effect of the skyrmions. The total Hall resistivity is given by the summation of the ordinary, topological, and anomalous resistivity, and expressed mathematically as

$$\rho_{xy} = \rho^O + \rho^A + \rho^T = R_O B + R_S \mu_0 M_Z + R_O P B_{em}^z \qquad (4)$$

where $R_O$ is the ordinary Hall coefficient, $B$ is the external magnetic field, $R_s$ is the anomalous Hall coefficient, $M_z$ is the magnetization component perpendicular to the film and $P$ is the spin polarization in the FM. The topological Hall resistance due to skyrmions emerges due to the emergent electromagnetic field seen by the conduction electrons. The conduction electrons couple with the local spin configuration of the skyrmion and acquire a Berry phase. The emergent magnetic field is given by [35]

$$B_Z^e = \frac{\Phi_Z^e}{A} = -\frac{h}{eA}\iint \frac{1}{4\pi} \mathbf{m} \cdot \left(\frac{\partial \mathbf{m}}{\partial x} \times \frac{\partial \mathbf{m}}{\partial y}\right) dxdy \qquad (5)$$

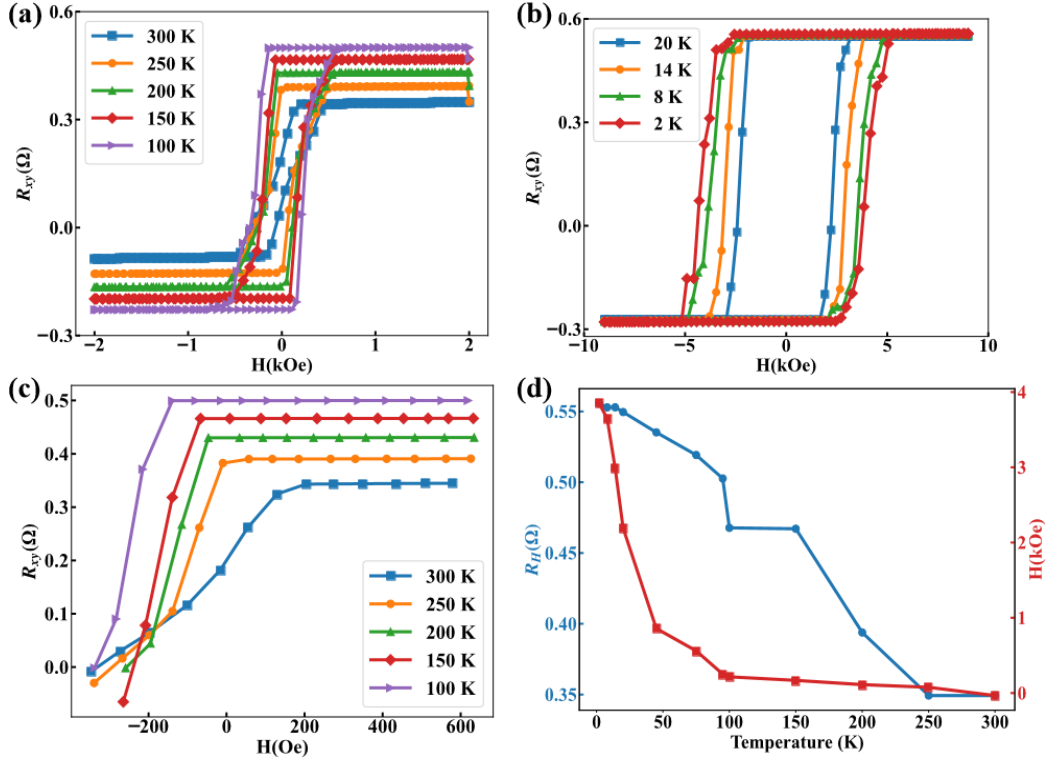

FIG. 3. Anomalous Hall resistance of $R_{xy}$ with H for different temperatures (a)100-300 K, (b)2-20 K. (c) The free layer switches more linearly at room temperature (indicating DW switching). (d) Saturating Anomalous Hall resistance and switching field as a function of temperature.

Which generates a topological resistivity

$$\rho_{xy}^T = PR_o \left|\frac{h}{e}\right| \frac{1}{A} \qquad (6)$$

Here, $P$ is the spin polarization of the conduction electrons, $R_o$ is the normal Hall coefficient, $h$ is the Planck's constant, e is the electron charge, A is the area of the cross-overlap and $\frac{h}{e}$ is the flux quantum.

To extract the topological resistivity from the measurement we first measured the normal Hall contribution as a linear fit to data at higher fields. The value of $R_O$ obtained is $1.44 \times 10^{-11}$ Ωm/T. The anomalous Hall effect completely dominates the Hall signal which is further proven by calculating the longitudinal conductivity $\sigma_{xx} \approx 0.5 \times 10^4 (\Omega cm^{-1})$, implying that the device is behaving at the edge of bad metal region [51].

Fig. 4(a) shows the temperature dependence of the anomalous Hall resistance and longitudinal resistance. We observe both resistances decreasing with increasing temperature which can be attributed to the weak localization or topological effects of skyrmions. In Fig. 4(b) we show the scaling of anomalous Hall resistivity concerning the longitudinal resistivity. The measured resistance values are taken at temperatures in the range of 2-300K and magnetic field 1 T. $R_{xy}$ behaves as the second-order polynomial as shown by the analytical fitting of the measured resistance.

$$\rho_{XY} = a(\rho_{XX0} + \rho_{XX0}^2 + \rho_{XXT}^2) + b\rho_{XXT} \qquad (7)$$

where, $a = -8.78 \times 10^{-5}$, $b = 0.1002$, and $\rho_{xx0}$ is the longitudinal resistivity resistance at 2K (approaching 0K). The anomalous Hall resistivity in general is believed to have three main physical constituents, intrinsic scattering, skew scattering, and side jump. The skew scattering due to impurity and phonon shows a linear relation with $\rho_{xx}$ while intrinsic contribution with its origin in the berry-phase curvature and side jump due to spin-orbit interaction show a quadratic behavior.

In literature [40], the scaling of anomalous Hall resistivity with longitudinal resistivity in Fe magnetic thin films showed the major contribution coming from intrinsic, side jump, and skew scattering. The skew scattering is predominantly dominated by residual resistivity with very negligible contribution coming from phonon-induced resistivity. So, there proposed scaling is mathematically given by [40]:

$$\rho_{XY} = \alpha \rho_{XX0} + \beta \rho_{XX0}^2 + b \rho_{XXT}^2 \qquad (8)$$

where $\alpha$ and $\beta$ are numerical constants.

Our measurements and analytical fitting reveal that the phonon contribution to the skew scattering term hence, anomalous Hall resistivity is significant. To extract the value of constants *a* and *b* we divided $R_{xy}$ by the $R_{xx}$ and fit it to a linear equation given below.

$$\frac{\rho_{XY}}{\rho_{XX}} = a\rho_{XX} + c \qquad (9)$$

The expected anomalous Hall resistivity depending only on the magnetization is given by

$$R_S M_Z \qquad (10)$$

Where $Rs = a(\rho_{XX0} + \rho_{XX0}^2 + \rho_{XXT}^2) + b\rho_{XXT} \qquad (11)$

To get a more realistic comparison between the expected anomalous Hall resistivity measured from the magnetization hysteresis. The squid VSM was performed on the Hall bar device which is more accurate when compared to other studies where measurements are performed on the whole magnetic samples. The measurements were carried out at different temperatures from 300K down to 2K. In comparison to the VSM measurements performed on the deposited samples, we observed the reduced saturation magnetization (by $\sim 10^2$) due to reduced magnetic moments in the Hall bar. Also,

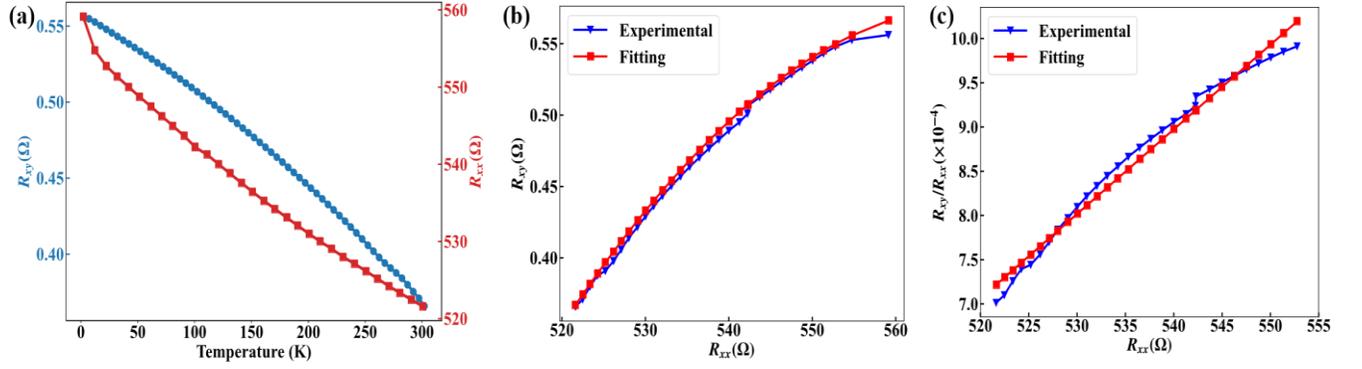

FIG. 4. (a) Anomalous Hall resistance and longitudinal resistance measured at temperatures 2 to 300K, H = 1 T. (b) The Anomalous Hall resistance as function of longitudinal resistance shows a quadratic behavior.
(c) The $R_{xy}/R_{xx}$ vs $R_{xx}$ is fitted as a linear equation to obtain the *a* and *b*

The magnetic saturation is more gradual as shown in Fig. 5 (a-c), indicating the switching by domain wall propagation and or skyrmion expansion/contraction. The topological effects on magnetization are more pronounced when measured for devices of small dimensions. Fig. 5(a-b) shows the measured anomalous Hall resistance hysteresis and extracted anomalous resistance for temperatures T= 300K and T= 250K. The extracted anomalous Hall resistance follows the behavior of the measured magnetization and is shown in Fig. 5 and Fig. 5(c) shows the magnetization and anomalous Hall resistance at 200K. The full anomalous Hall resistance hysteresis is provided in the supplementary material [Fig. S3]. In Fig. 5(b) we observe a minor deviation between the resistance curves corresponding to the extracted anomalous Hall resistance and the measured anomalous Hall resistance. A similar difference is noted in Fig. 5(a) and Fig.5 (c) between magnetization and resistance. This minor deviation is attributed to the topological resistance due to skyrmion nucleation in the magnetic field range of 0 Oe to 200 Oe.

The room temperature MFM images of the crossbar at 0 Oe, 50 Oe, 100 Oe, and 480 Oe are shown in Fig. 6. At 0 Oe the observed skyrmion density is low as seen in Fig. 6(a), more skyrmion is nucleated as magnetic field increases as shown for 50 Oe in Fig. 6(b). We observe the highest skyrmion density (both normal and distorted) around 50 Oe to 100 Oe. A few stripe domains are observed in this same field (see Fig. 6(c)). Finally, on increasing the magnetic field to 480 Oe the skyrmion density is reduced and magnetization saturation is observed but a few observed stripe domains remain. Note: The external magnetic field was applied via the self-made electromagnet with the capability of reaching $H_{max}$ = 490

Oe. The micromagnetic simulations at 0K shown in Fig. 8, show a similar magnetic texture evolution with the magnetic, as we observed in Fig. 6. (See supplementary video SV1).

We measured the topological resistivity for temperatures from 2K to 300K. Fig. 7(a-d) shows the variations of topological resistivity due to skyrmions throughout the magnetic hysteresis for temperatures 300K, 250K, 200K, and 50K. In all cases, the topological resistivity is maximum in the low field regime, and it starts decreasing with the increasing field. In particular, for the temperature 300K, the topological resistivity as a function of the magnetic field clearly shows a peak around 100 Oe and resistivity decreases with increasing magnetic field which indicates skyrmion annihilation. Thus, topological resistivity follows the trend seen in MFM and micromagnetic images as shown in Fig. 6 and SV1. At T=50 K the peak topological resistivity is 0.77 uΩcm and resistivity decreases to 0.3 uΩcm for 300K. When compared to 300K at low temperatures the possibility of stabilization of a large number of skyrmions is high. This is due to increased DMI, anisotropy, and saturation magnetization. The DMI scaling with temperature is found to be much stronger [41] as compared to the anisotropy and saturation magnetization.

In the presence of strong DMI and saturation magnetization $M_s$ the skyrmion density increases and we observe that measured topological resistance is highest at the 2K as shown in Fig. 8(b). The topological charge computed from micromagnetic simulations for a crossbar of dimensions (1024x 256) nm2, at different temperatures shows. At 0K the Q = -138, at 100K Q decreases to 127 for 200K the Q reduces to -66, and at 300K the Q stabilizes to a value of -10. Note: The small discrepancy between simulation and measured behavior is attributed to sample size differences and other interactions which can't be accommodated in the micromagnetic simulations. The decreasing skyrmion density/ topological charge at higher temperatures is shown in supplementary video SV2. To understand the relation between the topological resistivity and magnetic texture at 2K, the micromagnetic simulations of the similar scaled crossbar device at 2K are performed. As shown in Fig. 8 (b) the topological resistivity increases as the field is increased from 0 Oe to 3000 Oe and we measured the peak value around 3000 Oe. Like in the case of all the temperature measurements, the topological resistivity starts decreasing with increasing field and becomes zero at 9000 Oe. The micromagnetic simulations at 2K reveal that the topological charge increases with the field till 4200 Oe and it remains constant till 8800 Oe followed by an abrupt jump to zero. The topological charge being an integer (±1) is independent of the skyrmion size. This implies that the increasing magnetic field reduces the skyrmion size which changes the free layer magnetization as well. But the skyrmion density remains stable till the magnetic field 8800 Oe, this reflects in the topological charge profile as shown in Fig. 8(b). The topological resistivity due to skyrmions is given by

$$\rho_{xy}^T = P R_o \left|\frac{h}{e}\right| \frac{S}{A} \quad (8)$$

where *S/A* gives the skyrmion density as per Eq. (8) the topological resistivity should follow the topological charge characteristics and remain stable at the peak value till 8800 Oe. But as the magnetic field is increased above 3000 Oe the topological resistivity decreases as shown in Fig. 8(a) and Fig. 8(b). In micromagnetic simulations, we observe that the skyrmion size reduces as the magnetic field increases as shown in Fig. 8(d) also see supplementary video [Fig. S5]. We observe small skyrmions at 2K and the maximum skyrmion size obtained in the micromagnetic simulations is around 26 nm. A recent article on the effect of skyrmion size on topological conductivity in which the momentum (k) - space description of the system is used for the small skyrmions reveals that the skyrmion topological conductivity should decrease monotonically with decreasing skyrmion size. The topological resistivity is given by the ratio of topological conductivity and longitudinal conductivity. size results from a reduced number of bands in the magnetic Brillouin zone.

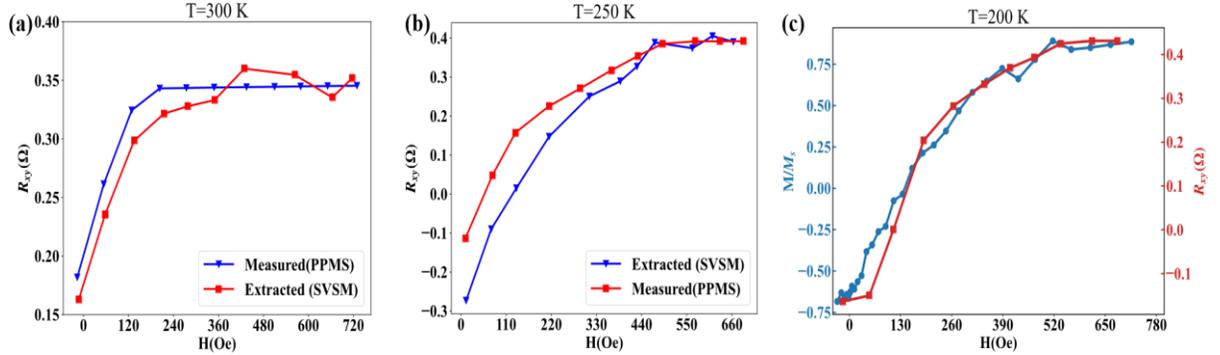

FIG. 5. (a) Anomalous Hall resistance and the normalized magnetization for T= 300K. (b) Anomalous Hall resistance and extracted anomalous Hall resistance for T= 300K. (c) Anomalous Hall resistance and extracted anomalous Hall resistance for T= 200K.

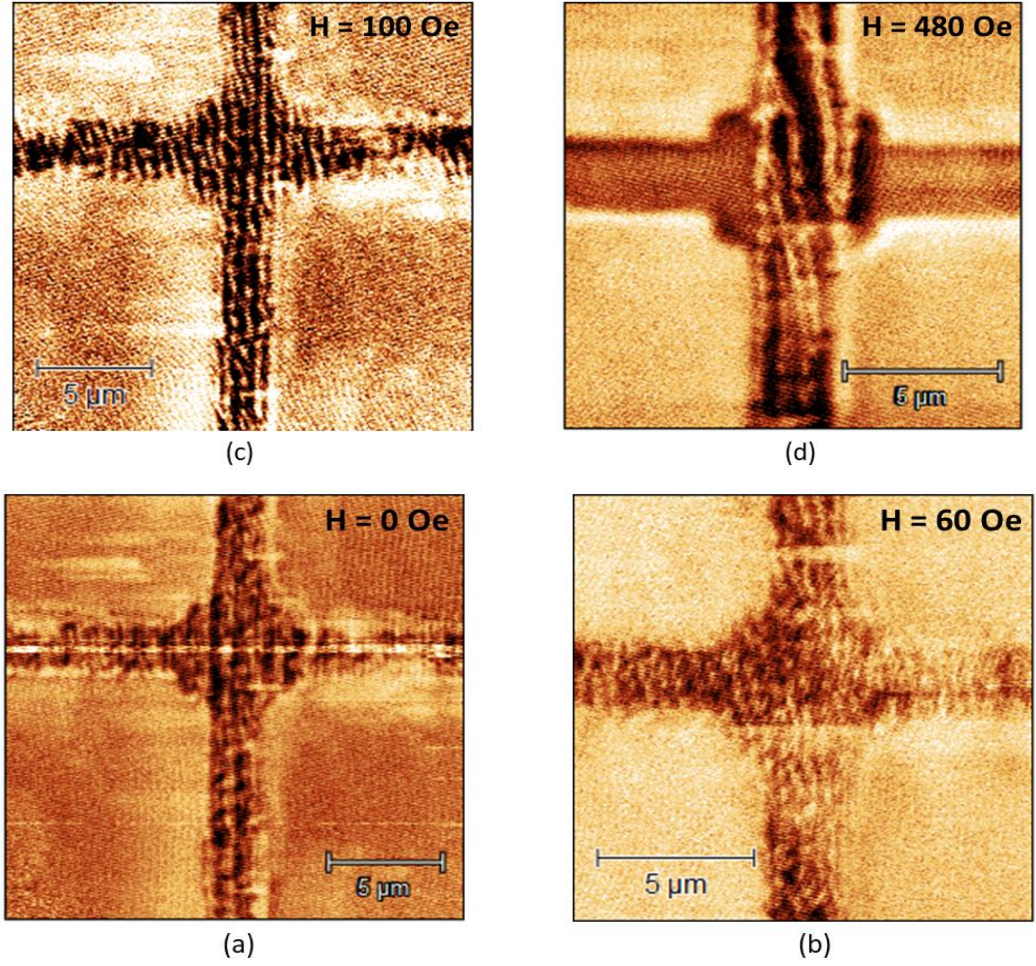

Fig. 6: Room temperature MFM scans at different magnetic fields. (a) H=0 Oe (a) H=50 Oe(a) H=100 Oe(a) H=480 Oe

$$\rho_{xy} \cong -\frac{\sigma_{xy}}{\sigma_{xx}} \qquad (9)$$

Thus, with decreasing skyrmion size the topological resistivity should also decrease as observed in Fig. 8(a) and Fig. 8(b). The decreasing resistivity with decreasing skyrmion

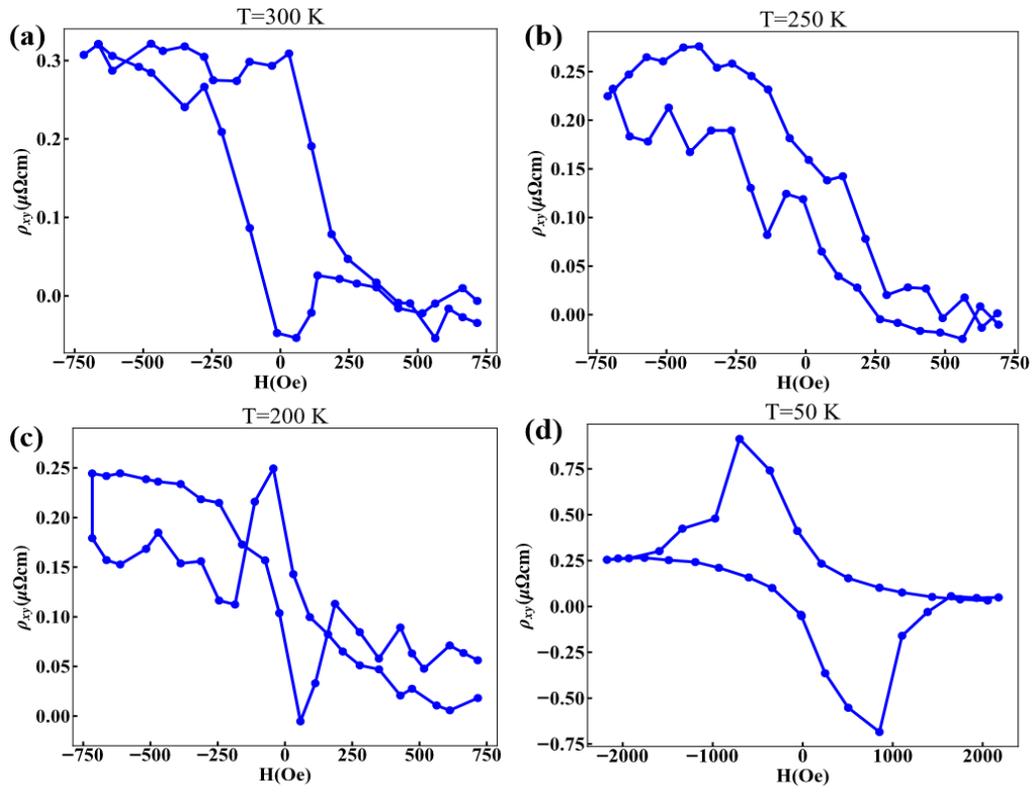

FIG. 7. Variation of Topological resistivity due to skyrmions throughout the magnetic hysteresis for temperatures (a) 300K, (b) 250K, (c) 200K, and (d) 50K.

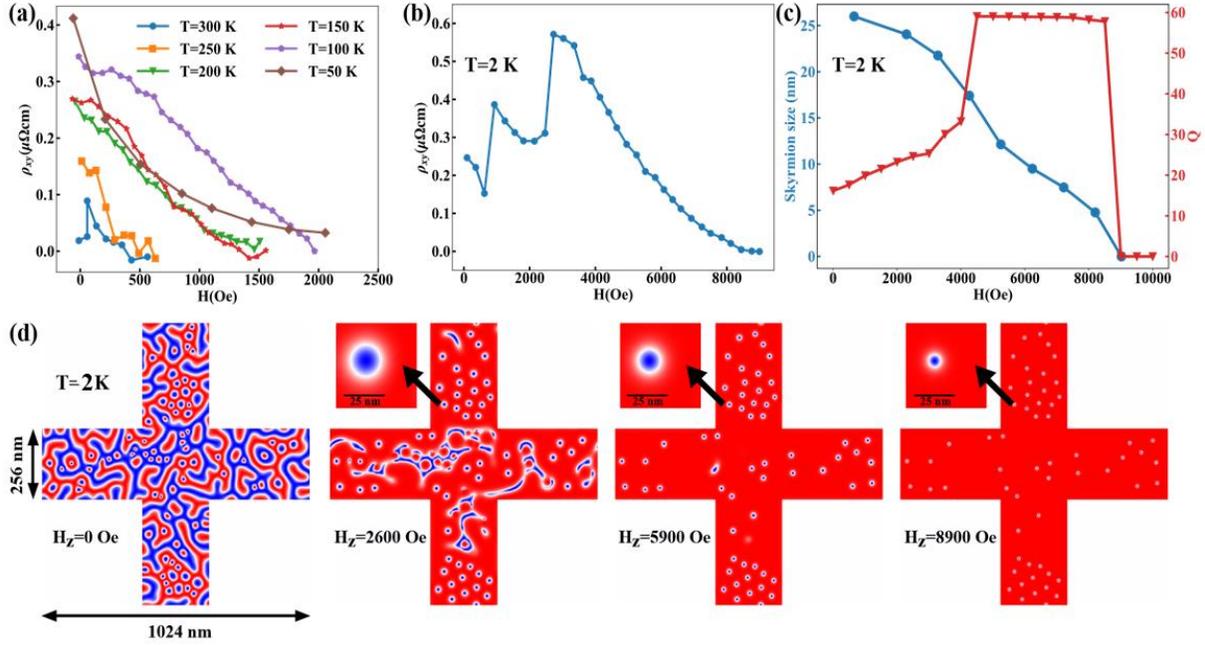

FIG. 8. a) Variation of $\rho_{xy}$ with the external magnetic field for temperatures T=300 K, 250 K, 200 K, 150 K, 100 K, and 50 K (b) Variation of $\rho_{xy}$, (c) average skyrmion size and topological charge ($Q$) with the external magnetic field at T=2 K. (d) Micromagnetic simulation results of the spin texture in a crossbar at the different magnetic field ($H_z$) performed at T=2 K.

## IV. SKYRMION-BASED NEURAL NETWORK IMPLEMENTATION

In Fig. 9 (a) we show the skyrmion device similar to the fabricated device. The area of the device considered in micromagnetic simulations is 1024 ×128 nm$^2$ while the thickness is (0.9 nm). The skyrmions-based synaptic device is divided into two regions a pre-synapse and an active region, as shown in Fig. 9(a). The skyrmions/domains are driven by SOT into the active region by passing a charge current from T-1 into T-2 and vice-versa. Depending upon the number of skyrmions/domains the magnetization behavior of the active region changes and this change in magnetization is read through terminals T-3 and T-4 as the anomalous Hall signal. We compute the magnetization from the MuMax and assign the corresponding resistance change as measured by the devices. The resistance potentiation and depression are demonstrated as shown in Fig. 9 (b). We observe a highly linear synaptic (minimal non-linearity≃ 0.05) behavior for both potentiation and depression. The device operation is visualized in supplementary video [SV2].

To address the device learning ability, we use the proposed synaptic device to train a 3-layer fully connected neural network FC-NN to classify the MNIST hand-written dataset (Fig. 9(c)). The network consists of three layers, including one input layer with 784 input neurons, a hidden layer with 100 neurons layer, and an output layer consisting of 10 neurons for 10-digit classification. The network is trained and tested on the MNIST dataset. We directly adopt the skyrmion-based synapse conductance and map these to the weights for network training. To train the network, the stochastic gradient descent algorithm was used during the backpropagation step. To benchmark this result, software-based perceptions in the same neural network architecture with ideal synaptic weights were used. After the algorithm terminates, the accuracy of our synapse device achieves approximately 89.5%, which is slightly slower than the software default 91.5% as shown in Fig. 9(d).

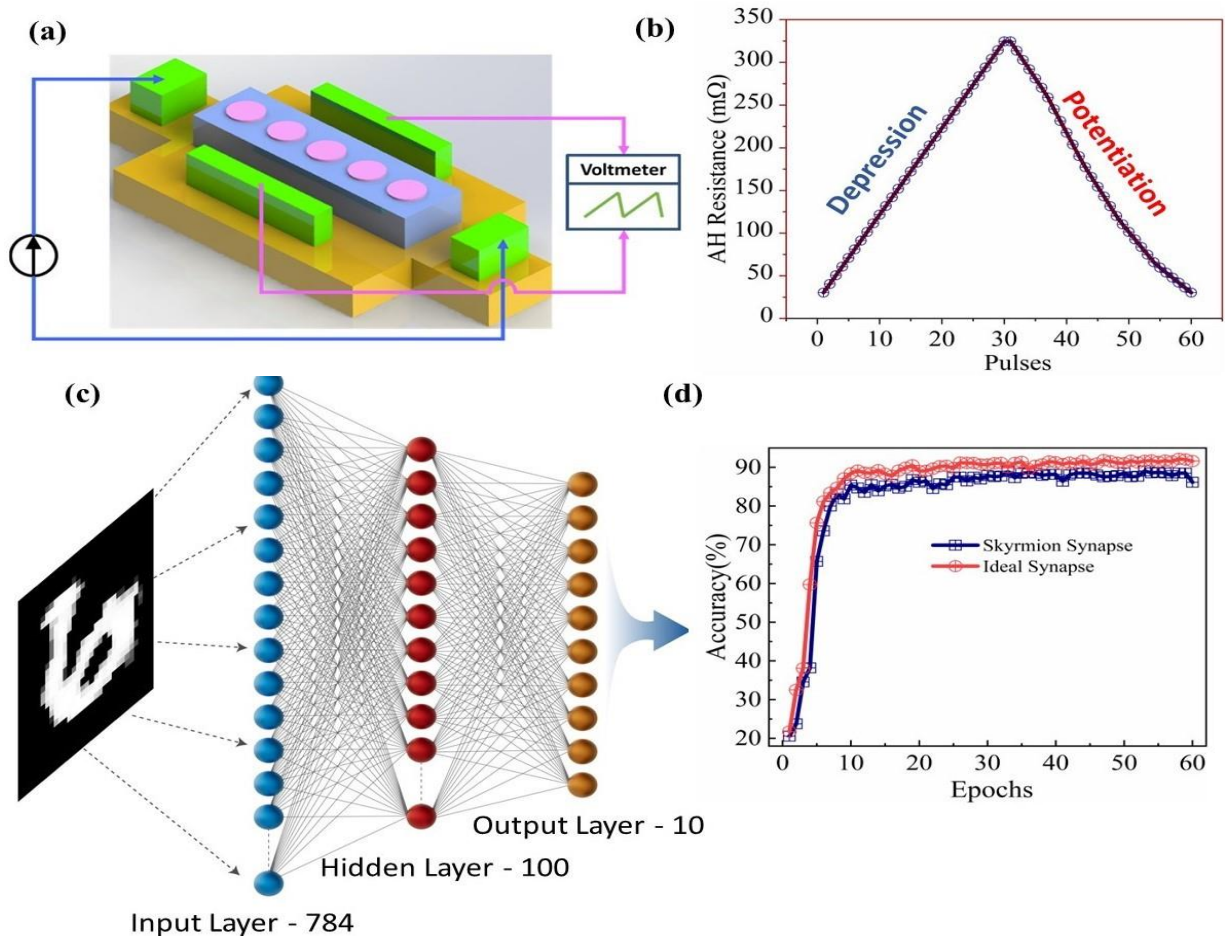

FIG. 9. a) Synaptic device design 1024x124 nm$^2$ SOT writing and AHE reading. (b) synaptic potentiation and depression show highly linear behavior. (c) 3-Layer fully connected neural network based on the proposed device. (d) Accuracy results achieved on MNIST data classification benchmarked with the ideal weights

This highlights that, beyond low power, the skyrmion-based device provides a significantly good linearity and learning ability that appears well suited to the characteristics of edge learning tasks and neuromorphic computing hardware.

**Conclusion**

In summary, we present the magnetic heterostructure Hall bar device and study the anomalous Hall and topological Hall signals in the device. The presented studies show a high resistance ON/OFF ratio devices where the anomalous Hall signals reading is utilized. The device has applications in both Hall bar sensors and memory especially for neuromorphic computing. The total anomalous Hall resistance is measured at temperatures from 2K to 300K and correlated with the expected anomalous Hall resistance measured from Squid-VSM magnetization data. Memory devices at lower temperatures have applications in cryogenic electronics. Furthermore, following the proper quantitative methodology, supported with MFM imaging and micromagnetic simulations. We show the topological Hall resistivity due to skyrmions, at different temperatures and magnetic fields. Motivated by the possibility of using anomalous and topological resistivity as an alternate reading mechanism. We propose a scaled version of the device as the synapse and map the magnetization evolution with the measured anomalous Hall resistance. The device shows a good linear potentiation/depression in the device with current pulses. When used in the simulated neural network architecture for MNIST data classification, we observe an accuracy of up to 90%, demonstrating the applications of the device for neuromorphic computing.

**Fabrication and Characterization**

The stacks were deposited by magnetron sputtering at room temperature onto 4-inch thermally oxidized Si wafers. The bottom layer on the substrate is 300nm silicon oxide by using thermal oxidization, then a multilayer stack is grown by using ultrahigh vacuum magnetron sputtering at room temperature in $5 \times 10^{-8}$ bar. The multilayer structure was, from the substrate side, A 300 nm SiO2 formed as an insulating layer, which is followed by growth of 10 or 15, or 20 stacks of Ta (5 nm)/CoFeB(x)/MgO (2 nm), and finally formed a stack of $SiO_2$(300nm)/[Ta (5 nm)/CoFeB(x)/MgO (2 nm)]$_{\times 15}$, where MgO layers were deposited by RF magnetron sputtering and the other layers were deposited by DC magnetron sputtering. The repeated deposition of Ta (5 nm)/CoFeB(x)/MgO (2 nm) was designed to increase skyrmion thermal stability and decrease the skyrmion size. The thickness of CoFeB is varied from 0.5nm to 1.1nm to find the optimum CoFeB thickness for the stabilization of room-temperature skyrmions. Then the deposited film was processed into micro-2.5x50 um$^2$ crossbars and followed by a lift-off process. To form electrical contacts. After the film deposition, we spin-coated AZ5214 photoresist of 1.6um thickness with hardened bake for 2 mins at 110 °C to use it in a positive tone. Then we patterned the crossbars on the resist using conventional photolithography. Ion beam etching using Ar gas to remove the exposed magnetic stacks outside the resist mask. During the etching, we monitored the conductivity of the etched region and stopped the etching when the signal from the 300-nm-thick silicon oxide layer appeared. To form electrical contacts, Ti (10 nm)/Au (100 nm) was deposited on the sides of the crossbars through sputtering at the rate of 0.8 nm/s. The electrical contact pads are defined by using photolithography and lifting off by immersion in acetone with ultrasonic processing for 5mins.

## Characterization and Imaging

The magnetic characterization of the samples for thickness optimization was done using normal vibration sample magnetometry (VSM) at room temperature. Samples with CoFeB thickness ranging from (0.5nm to 1.1 nm) were characterized. The deposition stack quality was confirmed with the cross-sectional TEM of the sample. After VSM we performed the imaging of the samples with multi-domain magnetic characteristics using magnetic force microscopy (MFM) based- Dimension Icon SPM. For probing we used CoIr coated MFM tip provided by Bruker Inc. The Hall measurements were done using a standard Hall measurement system Quantum Design PPMS capable of applying dc magnetic fields. We measured the crossbar devices at temperatures ranging from (2K to 300K) and at different magnetic fields depending upon the temperature. We furthermore, performed the superconductor-vibration sample magnetometry (Squid-VSM) on a 4mmx8mm sample carrying 4 crossbar devices. Using the magnetization measured from Squid-VSM we extracted the expected anomalous Hall resistivity. The MFM imaging of the devices in presence of a magnetic field and the current was done to relate the magnetic texture to the magnetization, and anomalous Hall resistivity.

II. MICROMAGNETICS

Magnetic skyrmions are described using their topological or skyrmion number Q, calculated as follows [26]:

$$Q = \frac{1}{4\pi} \int \int \boldsymbol{m} \cdot \left(\frac{\partial \boldsymbol{m}}{\partial x} \times \frac{\partial \boldsymbol{m}}{\partial y}\right) dx dy. \quad (1)$$

The spins projected on the xy-plane and normalized magnetization vector $\boldsymbol{m}$ can be determined by the radial function $\theta$, Vorticity $Q_v$ and helicity $Q_h$:

$$m(r) = [\sin(\theta)\cos(Q_v\varphi + Q_h), \sin(\theta)\sin(Q_v\varphi + Q_h), \cos(\theta)]. \quad (2)$$

The vorticity number is related to the skyrmion number as follows [4]:

$$Q = \frac{Q_v}{2}\left[\lim_{r \to \infty} \cos(\theta(r)) - \cos(\theta(0))\right]. \quad (3)$$

Micromagnetic simulations were performed using MuMax having the Landau–Lipschitz–Gilbert (LLG) equation as the basic magnetization dynamics computing unit[51]. The LLG equation describes the magnetization evolution as follows:

$$\frac{d\hat{m}}{dt} = \frac{-\gamma}{1+\alpha^2}\left[\boldsymbol{m} \times \boldsymbol{H}_{eff} + \boldsymbol{m} \times (\boldsymbol{m} \times \boldsymbol{H}_{eff})\right] \quad (4)$$

where $\boldsymbol{m}$ is the normalized magnetization vector, $\gamma$ is the gyromagnetic ratio, $\alpha$ is the Gilbert damping coefficient, and

$$\boldsymbol{H}_{eff} = \frac{-1}{\mu_0 M_S}\frac{\delta E}{\delta m} \quad (5)$$

is the effective MF around which the magnetization process occurs. The total magnetic energy of the free layer includes exchange, Zeeman, uniaxial anisotropy, demagnetization, and DMI energies [52,53].

$$E(\boldsymbol{m}) = \int_V [A(\nabla \boldsymbol{m})^2 - \mu_0 \boldsymbol{m} \cdot \boldsymbol{H}_{ext} - \frac{\mu_0}{2}\boldsymbol{m} \cdot \boldsymbol{H}_d - K_u(\hat{u} \cdot \boldsymbol{m}) + \varepsilon_{DM}] dv \quad (6)$$

where A is the exchange stiffness, $\mu_0$ is the permeability, Ku is the anisotropy energy density, $H_d$ is the demagnetization field, and $H_{ext}$ is the external field; moreover, the DMI energy density is then computed as follows:

$$\varepsilon_{DM} = D[m_z(\nabla \cdot \boldsymbol{m}) - (\boldsymbol{m} \cdot \nabla) \cdot \boldsymbol{m}] \quad (7)$$

The spin–orbit torque is then added in the form of modified STT in MuMax [48].

$$\boldsymbol{\tau}_{SOT} = -\frac{\gamma}{1+\alpha^2} a_J[(1 + \xi\alpha)\boldsymbol{m} \times (\boldsymbol{m} \times \boldsymbol{p}) + (\xi - \alpha)(\boldsymbol{m} \times \boldsymbol{p})] \quad (8)$$

$$a_J = \left|\frac{\hbar}{2M_se\mu_0}\frac{\theta_{SH}j}{d}\right| \quad \text{and} \quad \boldsymbol{p} = sign(\theta_{SH})\boldsymbol{j} \times \boldsymbol{n}$$

where $\theta_{SH}$ is the spin Hall coefficient of the material, $j$ is the current density, and d is the free layer thickness.

**Author Contribution:** A. H. Lone conceived the idea and fabricated the devices along with X. Zou. A. H. Lone did the characterization of the devices and performed the micromagnetic simulations. The paper was written by A. H. Lone with support from D. Das, K. Fong, G. Setti and H. Fariborzi. The project was supervised by H. Fariborzi and G. Setti.

- **References**